\documentclass[twocolumn, twocolappendix]{aastex63}

\usepackage{CJKutf8}
\usepackage[dvipsnames]{xcolor}
\usepackage{lipsum}
\usepackage[hyphens]{url}
\usepackage{hyperref}
\usepackage[hyphenbreaks]{breakurl}
\usepackage{xspace}

\usepackage{tabularx}
\newcommand{\colA}{0.15\textwidth}
\newcommand{\colB}{0.12\textwidth}
\newcommand{\colC}{0.25\textwidth}
\newcommand{\colD}{0.17\textwidth}
\newcommand{\colE}{0.17\textwidth}

\newcommand{\yes}{\textcolor{ForestGreen}{\checkmark}}
\newcommand{\maybe}{$\textcolor{Dandelion}{\triangle}$}
\newcommand{\no}{$\textcolor{BrickRed}{\times}$}

\definecolor{michelle}{RGB}{0, 114, 178}    
\definecolor{jo}{RGB}{213, 94, 0}           
\definecolor{kartheik}{RGB}{0, 158, 115}    
\definecolor{kelly}{RGB}{204, 121, 167}     
\definecolor{jan}{RGB}{140, 86, 75}         
\definecolor{josh}{RGB}{120, 55, 123}         
\definecolor{sarah}{RGB}{51, 204, 204}         
\definecolor{jeff}{RGB}{199, 167, 227}      
\definecolor{ashish}{RGB}{209, 207, 27}      

\newcommand*\todopar[1]{{\color{blue}A paragraph that #1~.} {\color{gray}\lipsum[1]}}
\renewcommand*\todopar[1]{}

\newcommand*\apprentice{\textit{Craftsmanship}\xspace}
\newcommand*\gettherefast{\textit{High Leverage}\xspace}
\newcommand*\stewardship{\textit{Data Stewardship}\xspace}
\newcommand*\qualityaboveall{\textit{Trustworthiness}\xspace}

\shorttitle{AI Policy}
\shortauthors{Ntampaka et al.}

\begin{document}

\correspondingauthor{Michelle Ntampaka}
\email{mntampaka@stsci.edu}

\author[0000-0002-0144-387X]{Michelle Ntampaka}
\affiliation{Space Telescope Science Institute, Baltimore, MD 21218,USA}
\affiliation{Department of Physics \& Astronomy, Johns Hopkins University, Baltimore, MD 21218,USA}

\author[0000-0003-4052-7838]{S. Burke-Spolaor}
\affiliation{Department of Physics and Astronomy, West Virginia University, Morgantown, WV 26506, USA}
\affiliation{Center for Gravitational Waves \& Cosmology, West Virginia University, Morgantown, WV 26505, USA}
\affiliation{Department of Physics \& Astronomy, Johns Hopkins University, Baltimore, MD 21218,USA}

\author[0000-0002-0104-3593]{Ioana Ciuc\u{a}}
\affiliation{Stanford University, Stanford, CA 94305, USA}

\author[0009-0000-2927-2104]{Ana Maria Delgado}
\affiliation{Department of Physics \& Astronomy, Johns Hopkins University, Baltimore, MD 21218,USA}

\author[0000-0001-9298-3523]{Kartheik G. Iyer}
\affiliation{Center for Computational Astrophysics, Flatiron Institute, New York, NY 10010, USA}

\author[0000-0002-8130-1440]{Kelly Lockhart}
\affiliation{Center for Astrophysics $|$ Harvard \& Smithsonian, Cambridge, MA 02138, USA}

\author[0000-0002-9912-5705]{Adele Plunkett}
\affiliation{National Radio Astronomy Observatory, Charlottesville, VA 22903, USA}

\author[0000-0003-3204-8183]{Mercedes López-Morales}
\affiliation{Space Telescope Science Institute, Baltimore, MD 21218,USA}

\author[0000-0003-2242-0244]{Ashish~A.~Mahabal}
\affiliation{Division of Physics, Mathematics and Astronomy, California Institute of Technology, Pasadena, CA 91125, USA}
\affiliation{Center for Data Driven Discovery, California Institute of Technology, Pasadena, CA 91125, USA}

\author[0000-0003-3204-8183]{Susan E.~Mullally}
\affiliation{Space Telescope Science Institute, Baltimore, MD 21218,USA}

\author[0009-0000-0170-6302]{Rohit Raj}
\affiliation{Center for Astrophysics $|$ Harvard \& Smithsonian, Cambridge, MA 02138, USA}

\author[0000-0001-6558-5183]{Jan Reerink}
\affiliation{European Space Agency (ESA), European Space Astronomy Centre (ESAC), Madrid, Spain}

\author[0000-0002-6148-7903]{Jeffrey Smith}
\affiliation{SETI Institute, Mountain View, CA 94043, USA}

\author[0000-0003-2573-9832]{Joshua S. Speagle (\begin{CJK*}{UTF8}{gbsn}沈佳士\ignorespacesafterend\end{CJK*})}
\affiliation{Department of Statistical Sciences, University of Toronto, 9th Floor, Ontario Power Building, 700 University Avenue, Toronto, ON M5G 1Z5, Canada}
\affiliation{David A. Dunlap Department of Astronomy \& Astrophysics, University of Toronto, 50 St George Street, Toronto, ON M5S 3H4, Canada}
\affiliation{Dunlap Institute for Astronomy \& Astrophysics, University of Toronto, 50 St George Street, Toronto, ON M5S 3H4, Canada}
\affiliation{Data Sciences Institute, University of Toronto, 17th Floor, Ontario Power Building, 700 University Avenue, Toronto, ON M5G 1Z5, Canada}

\author[0000-0002-0104-3593]{John Soltis}
\affiliation{Space Telescope Science Institute, Baltimore, MD 21218,USA}

\author[0000-0002-5077-881X]{John~F.~Wu}
\affiliation{Space Telescope Science Institute, Baltimore, MD 21218,USA}
\affiliation{Department of Physics \& Astronomy, Johns Hopkins University, Baltimore, MD 21218,USA}
\affiliation{Department of Computer Science, Johns Hopkins University, Baltimore, MD 21218, USA}

\author[0000-0002-9851-2850]{Mikaeel Yunus}
\affiliation{Department of Physics \& Astronomy, Johns Hopkins University, Baltimore, MD 21218,USA}

\title{How to Craft the Right Language AI Policy For Your Research Group\\ (Some Assembly Required)\\}

\begin{abstract}
Language AI is rapidly becoming part of the astronomy research ecosystem, prompting research teams to develop policies governing its use.  But resources and advice for AI adoption assume that all research groups share the same goals and values. 
This paper lays out an argument for why there is no single ``correct'' AI policy for astronomy research groups. Instead, we introduce four research laboratory archetypes with competing research priorities, and we use them to explore how a small laboratory or research group's priorities shape decisions about AI's impact on research productivity, scientist development, scientific integrity, and data governance.  Rather than prescribing a universal set of rules, this paper provides a framework for aligning AI policies with a group's scientific values and mission. The objective is to help research leaders decide how Language AI should be used within their particular research environment.\vspace{2em}
\end{abstract}

\section{Introduction}
\label{sec:introduction}


Language AI is a double-edged sword that can simultaneously make an astronomy researcher more productive while eroding expertise.  Because AI tools can perform tasks such as drafting and revising text, summarizing documents, answering questions, translating between languages, writing code, and assisting with information retrieval and analysis, they accelerate many aspects of scientific research, but this speedup comes with a cost.
For example, a student who uses AI to write code may complete a project more quickly while learning less about software development, and a senior scientist who uses AI to summarize literature may absorb information more efficiently while becoming less practiced in the critical evaluation and synthesis of scientific ideas. The same tool that accelerates research can also alter how expertise is developed, evaluated, and maintained.

Astronomy is one field among many where Language AI has emerged rapidly, shifting the way we research.  The Astro2020 Decadal Survey  \citep[\textit{Pathways to Discovery},][]{decadal2020} identified the crucial role that artificial intelligence and machine learning could play in this next decade: ``Machine learning has already shown significant success at providing tools for identifying anomalies in data, and can speed up parameter estimation in large data sets by significant factors\ldots  These techniques could lead to transformative discoveries from the new data sets available in the 2020s.''
In the short time since Astro2020's publication, Large Language Models (LLMs) have radically changed the way researchers interact with information. LLMs represent a rapidly emerging class of artificial intelligence systems capable of processing and generating natural language. Modern LLMs are typically built using transformer architectures \citep{vaswani2017attention} and trained on extremely large corpora of text and code.  LLM-based systems operate as general-purpose assistants capable of supporting a wide range of knowledge work \citep{GPT3}, including literature review, brainstorming, drafting text, and translating conceptual ideas into executable code \citep[e.g.,][]{codegen, copilot}. 

Language AI tools based on these technologies are rapidly becoming part of the research ecosystem, prompting many research teams to develop policies governing their use. However, recommendations for adopting AI often assume that all research groups share the same goals and values. In practice, a laboratory racing toward the next discovery, a group focused on training students, a team prioritizing methodological rigor, and a lab serving data and software may each have legitimate reasons to engage with AI in very different ways. As a result, a policy that is appropriate for one research environment may be ineffective or even counterproductive in another.

The empirical literature on AI in knowledge work supports both optimism and unease, often within the same study. \cite{noy2023experimental} found that professionals with weaker baseline skills gained more from ChatGPT than strong writers, compressing the productivity distribution. This is the leveling effect behind the AI-as-equalizer narrative. However, \cite{delllacqua2023jagged} found that workers using GPT-4 performed markedly better on tasks within the model's competence and markedly worse on tasks just outside of it, and most were unable to tell the difference. Astronomy does not start from neutral conditions. Who is read and who is cited is already structured by inequities the field has measured. For example, papers authored by women receive $10.4 \pm 0.9\%$ fewer citations than papers with identical non-gender-specific properties authored by men \citep{caplar2017}. Because generative models are fundamentally shaped by their training data, models fine-tuned on the astronomical literature may inherit its existing biases, such as a preference for high-citation institutions, majority-language authors, and mainstream methodologies.

In Astronomy, Language AI tools are increasingly being explored to accelerate every stage of scientific discovery, from hypothesis generation to proposal writing, from developing analysis code to drafting manuscripts \citep[e.\,g.][]{moss+aicosmologist}. At the same time, researchers are developing astronomy-specific benchmarks and domain-adapted language models to evaluate and improve the performance of these models on astrophysical tasks \citep[e.g.,][]{2025arXiv250520538J, 2026arXiv260318145S}. These developments suggest that Language AI systems may become an increasingly important component of the astronomical research workflow, acting as interfaces between human scientific reasoning and computational infrastructure, while raising new questions about authorship, reliability, and responsible use within the scientific process.

But learning and teaching are also core elements of the scientific process. A metastudy by \cite{wang2025effect} found that ChatGPT had an overall positive impact on learning performance, but not all findings are positive. For example, \cite{2025arXiv250608872K} suggests decreased cognitive engagement when writing is assisted with Language AI tools like ChatGPT, highlighting that the convenience of the tools comes with cognitive costs. \cite{2026arXiv260329039H} concludes that AI does not improve outputs across the board in Astronomy, but rather its effects are strongly dependent on the task attempted, the policy applied, and the language model used. \cite{Liu2026} shows that, while AI assistance improves short-term performance across several tasks, people perform significantly worse when the AI is taken away, and they are far more likely to give up.

As Language AI tools become increasingly integrated into research workflows, individual research groups should decide how to weigh the benefits and potential drawbacks of these tools. Establishing a group-level AI policy is a way to codify and articulate shared expectations and norms before problems arise. Creating a laboratory handbook for small research teams \citep[as described in][]{mehr2020write} can save time, curate knowledge, and, most importantly, provide trainees with clarity about how they should work; it can be a cultural glue that aligns the team with best practices.

Professional societies in astronomy have recognized the importance of explicitly stating certain norms.  The American Astronomical Society (AAS), for example, maintains a Code of Ethics \citep{AASCodeEthics} that captures shared professional standards on a range of topics, including bullying, authorship, and conflicts of interest. A research group's AI policy is a more granular extension of these community principles that extends field-wide expectations into concrete guidance for how a specific team chooses to conduct its research.  Although these terms can have distinct meanings in some organizational contexts, we use ``research lab,'' ``research team,'' and ``research group'' interchangeably throughout this paper to refer to a small team of researchers working toward common scientific goals.

Rather than just being a set of rules, the policy of a research lab is a way to align the team's values and practices.  For such a policy to work, we recommend that you create the Language AI policy together and view it as a living document that will evolve with AI's changing capabilities. No two research groups are the same, and there is no one-size-fits-all policy that will meet the entire community's needs. But the process of explicitly defining the data culture a research group aims to create, and then clarifying and communicating expectations when engaging with Language AI, will set the team up for success in the rapidly evolving AI landscape.

Rather than prescribing a single rigid standard, our goal in this white paper is to articulate the key considerations and trade-offs that research groups may wish to weigh when developing their own policies. In this sense, the document is intended as a starting point for discussion: by describing several laboratory archetypes (Section \ref{sec:fourlabs}) and the principal axes along which decisions should be made (Sections \ref{sec:productivity}-\ref{sec:data}), we hope to facilitate your research team's conversations about how these tools should be incorporated into their scientific practice.  We lay out a discussion about how each lab archetype might adopt or prohibit the use of specific AI tools (Section \ref{sec:sample}) and we discuss the inherent tensions and tradeoffs in any Language AI technology adoption (Section \ref{sec:conclusion}).  

Appendix \ref{sec:example} presents the AI policy developed and adopted by one of the research groups represented among this paper's authors, but your research group's policy comes with ``some assembly required.''  We do not advocate for the sample policy to be adopted as-is, but rather, we present it as a concrete example of how one laboratory translated its particular values, priorities, and constraints into an operational AI policy.  Readers are encouraged to use the blank worksheet in Appendix \ref{sec:blank} to facilitate discussions as a starting point for building a lab AI policy that fits the priorities of your team.

\begin{figure*}
    \hspace{3.7cm}
    \includegraphics[width=0.75\linewidth]{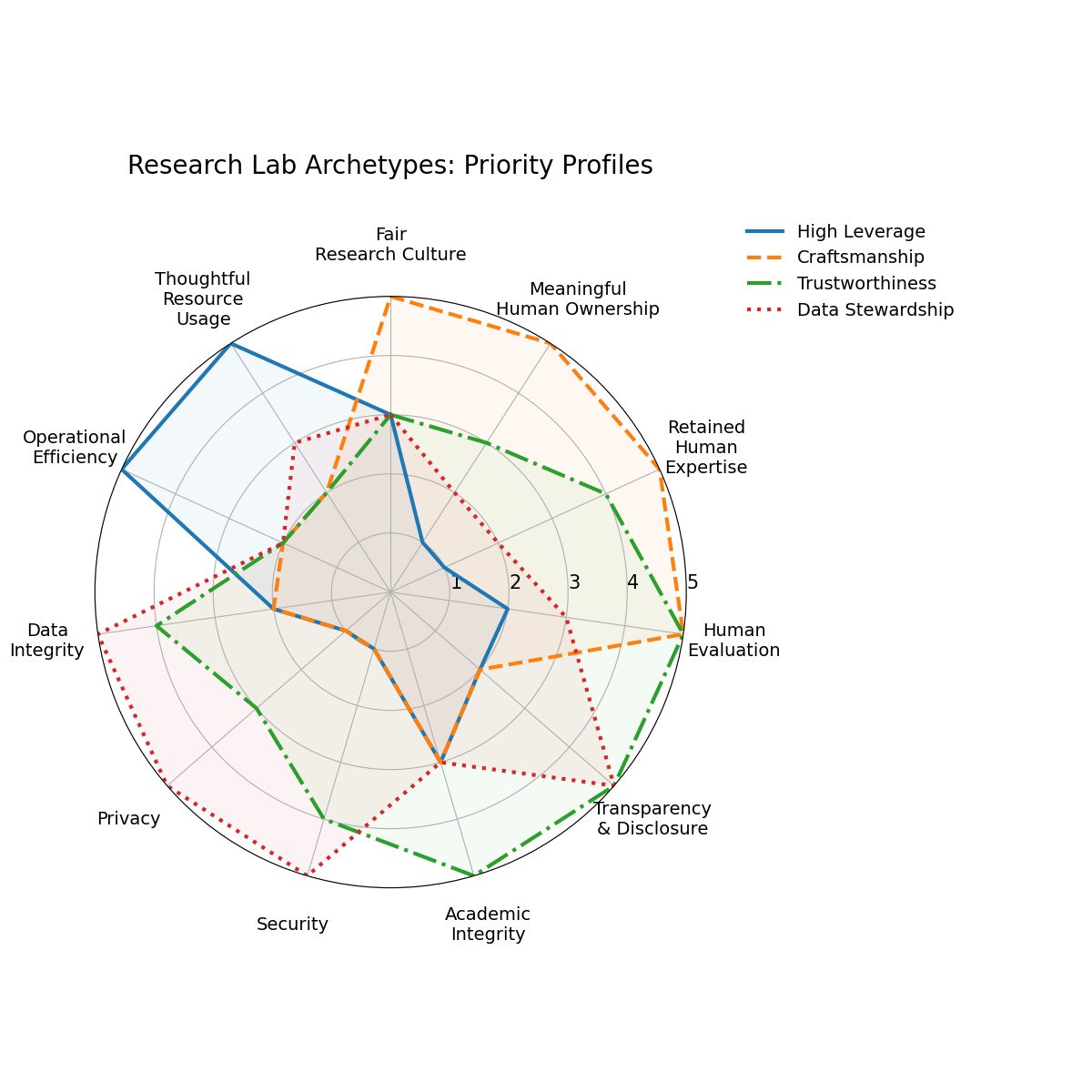}
    \caption{Radar diagram for visualizing how each of the laboratory archetypes might prioritize eleven objectives, where a value of 5 indicates that a metric is highly prioritized by a given archetype, and a value of 1 indicates a low priority.  The \gettherefast Lab Archetype (blue solid) prioritizes elements of Research Productivity (\S\ref{sec:productivity}), such as operational efficiency and thoughtful resource usage.
    The \apprentice Lab Archetype (orange dashed) prioritizes elements of Scientist Development (\S\ref{sec:human}), such as fair research culture, meaningful human ownership, and retained human expertise.
    The \qualityaboveall Lab Archetype (green dash dotted) prioritizes elements of Scientific Integrity (\S\ref{sec:integrity}), such as human evaluation, transparency \& disclosure, and academic integrity.
    The \stewardship Lab Archetype (red dotted) prioritizes elements of Data Governance (\S\ref{sec:data}), such as security, privacy, and data integrity.  A blank version of this diagram is available in Appendix \ref{sec:blank} for mapping out your own research priorities.}
    \label{fig:spider}
\end{figure*}

\bigskip
\section{four lab archetypes}
\label{sec:fourlabs}

\begin{table*}[t]
\caption{Illustrative research laboratory archetypes and their approaches to generative AI.}
\label{tab:archetypes}
\begin{center}
\begin{tabular}{||l|l|l|l|l||}
\hline
\hline
\parbox[t]{\colA}{\textbf{Archetype}} &
\parbox[t]{\colB}{\textbf{Optimizes For}} &
\parbox[t]{\colC}{\textbf{Key Question}} &
\parbox[t]{\colD}{\textbf{AI's Primary Role}} &
\parbox[t]{\colE}{\textbf{Most Willing\\ to Trade Away}} \\
\hline
\hline

\parbox[t]{\colA}{\gettherefast} &
\parbox[t]{\colB}{Throughput \& Efficiency \\} &
\parbox[t]{\colC}{Does this help us accomplish our goals efficiently?} &
\parbox[t]{\colD}{Force Multiplier} &
\parbox[t]{\colE}{Training \& Oversight} \\
\hline

\parbox[t]{\colA}{\apprentice} &
\parbox[t]{\colB}{Expertise\\ Development\\} &
\parbox[t]{\colC}{Are we developing capable\\ researchers?} &
\parbox[t]{\colD}{Tutor} &
\parbox[t]{\colE}{Efficiency\\} \\
\hline

\parbox[t]{\colA}{\qualityaboveall} &
\parbox[t]{\colB}{Reliability\\ \& Rigor\\} &
\parbox[t]{\colC}{Can I trust the process and\\ result?} &
\parbox[t]{\colD}{Verified Assistant} &
\parbox[t]{\colE}{Speed} \\
\hline

\parbox[t]{\colA}{\stewardship} &
\parbox[t]{\colB}{Security \&\\ Responsibility\\} &
\parbox[t]{\colC}{Is this approach secure,\\ compliant, and responsible?} &
\parbox[t]{\colD}{Controlled Tool} &
\parbox[t]{\colE}{Convenience\\} \\

\hline
\hline
\end{tabular}
\end{center}
\end{table*}

Before deciding on a Language AI policy, you must first decide what kind of research team you are, and what kind of team you want to be.  In this section, we describe four research laboratory archetypes that capture common philosophies of scientific practice in research groups. These archetypes are intentionally exaggerated caricatures\textemdash{}no laboratory will fit perfectly within a single category, but is more likely to be a mix of two or more.  These archetypes provide a conceptual framework for examining how differing research priorities may shape decisions regarding the adoption, governance, and use of Language AI.  Figure \ref{fig:spider} illustrates how each archetype might prioritize different research priorities; archetypes are described below, and the eleven research priorities are described in more detail in Sections \ref{sec:productivity}-\ref{sec:data}.

The \gettherefast Lab Archetype will primarily use AI as a force multiplier or a tool to remove bottlenecks that slow progress. Labs that focus on speed and being the first to a new discovery are one way to be a \gettherefast laboratory archetype, but these labs may have a broader definition; rather than maximizing papers and minimizing time, this archetype maximizes scientific impact within limited resources.

The \apprentice Lab Archetype works under the philosophy that the primary goal of research is not papers, but scientists. This research lab prioritizes deep expertise that is developed through practice and failure. Such a research lab will primarily use AI as a tutor, helping students learn while avoiding shortcuts that bypass the development of expertise.

The \qualityaboveall Lab Archetype values reproducibility, validation, transparency, and methodological rigor. This research lab will view AI with skepticism and will prioritize independent checks of AI-generated analyses. The team will not adopt AI hastily, but will integrate it in ways that preserve the highest standards of scientific integrity.

The \stewardship Lab Archetype views itself as a custodian of scientific data, software, institutional resources, and public trust. This laboratory prioritizes responsible data management, security, privacy, compliance, and long-term sustainability. AI tools are adopted by this archetype when they can be used safely and responsibly, but researchers are expected to protect the integrity of the data and systems entrusted to them.

These archetypes are caricatures contrasting philosophies rather than representing the full complexity of real research groups.  They are not intended to represent all possible research cultures, and most research groups will incorporate elements of multiple philosophies and adopt a more balanced approach that seeks to advance a subset of these priorities simultaneously, rather than leaning heavily into growth in a single area.  However, the archetypes provide a useful framework for understanding how differing priorities can lead to distinct approaches to Language AI adoption. By identifying the values and objectives that most strongly characterize their research group, its leaders can develop AI policies that are aligned with the scientific mission, training goals, and standards of rigor of the team rather than adopting a one-size-fits-all approach.

Figure \ref{fig:spider} is a conceptual framework showing how each of the laboratory archetypes may prioritize their efforts. We note that the authors of this white paper do not all ascribe to the same research philosophy, nor do we share identical views on where the appropriate boundaries should lie for the use of Language AI tools in research. The authors themselves span multiple archetypes, holding differing views regarding the appropriate role of Language AI. 
We see this as a good thing, because the discussions that motivated this paper arose from those differences.  Scientists are a diverse group of people with different workflows, goals, and priorities.  And so it follows that reasonable researchers will draw different conclusions about what constitutes appropriate or inappropriate use of Language AI in their specific contexts.  In the next sections, we expand on the main decision factors for each of these laboratory archetypes.

\bigskip
\section{Research Productivity}
\label{sec:productivity}
Research productivity is the highest priority for the \gettherefast research laboratories, which focus on maximizing research excellence within the available resources.  They are interested in adopting tools that address their key concern: \textbf{Does this help us accomplish our goals efficiently?}  

These laboratories tend to view generative AI as part of the natural progression of increasingly powerful scientific tools, the next logical step in a progression that introduced computers, numerical methods, and machine learning to the field in the past. They place greater value on advancing scientific knowledge than on preserving a uniquely human role in every step of the research process, arguing that the benefits of discovery are shared broadly regardless of the specific tools used to achieve it.  They are most likely to look for places where AI can reduce repetitive tasks or remove friction from the research process. Rather than spending researcher time on low-value implementation details (e.g., adjusting plot and paper formatting or debugging routine programming errors), this research group will opt to adopt tools that allow them to devote more of their time to designing experiments, interpreting results, and refining scientific ideas.  This research group recognizes that AI can accelerate the path from concept to scientific insight by reducing the effort required to translate ideas into polished figures, publication-ready manuscripts, and working code.

\subsection{Operational Efficiency}
Operational efficiency includes reducing bottlenecks, eliminating repetitive work, streamlining research workflows, and directing limited resources toward activities that generate the greatest scientific return. Scientific output may take many forms, including publications, analyses, completed projects, successful grant proposals, software and data products, community resources, career advancement, scientific influence, and the ability to respond quickly to time-sensitive opportunities. From this perspective, productivity is not simply a measure of papers produced per year, but of the total scientific value generated.

\subsection{Thoughtful Resource Usage}
Thoughtful resource use considers whether limited resources are being directed toward activities that create the greatest scientific return. Research groups consume many kinds of resources, including researcher time, clock time, funding, personnel, computational power, equipment, subscriptions, administrative effort, attention, and cognitive load. Labs also incur less visible costs, including environmental impact, risk exposure, and opportunity cost.  

\subsection{Policy Considerations}
AI is not universally a productivity enhancer. Depending on the task, current models may waste more time and resources than they save.  Effective AI use is itself a skill, requiring practice to recognize when AI is likely to accelerate a task and when it is more likely to introduce unnecessary work.  AI tools contribute to research productivity not only when they result in faster work, but also when they reduce the total cost of producing meaningful scientific outcomes by automating routine tasks, accelerating analysis, minimizing the effort required to develop software and documentation, and allowing researchers to explore ideas more rapidly.  When considering research productivity, consider whether the tool helps you to meet your professional goals efficiently.  For laboratories that prioritize research productivity, the relevant question is not simply whether AI can make a task faster, but whether it produces a good return on investment.


\bigskip
\section{Scientist Development}
\label{sec:human}
Scientist Development is the highest priority of the \apprentice laboratories.  These teams value mentorship and growth opportunities, and their AI adoption addresses the core question: \textbf{Are we developing capable researchers?}  

These laboratories focus on supporting three key areas:  fostering a healthy and sustainable research culture, maintaining meaningful human ownership of research, and retaining the human expertise needed to carry out the lab's scientific research.

\subsection{Fair Research Culture}
\label{sec:culture}
The fairness questions raised by AI are usually framed as questions of access: who can afford the better model or tool. A second risk is one of \textit{enforcement}. As the field develops norms around acceptable AI use, the cost of those norms does not fall evenly, and consequences of undisclosed AI usage might vary across career stages, positions, and institutions.  Those who write the emerging rules might bear the smallest cost when those rules are wrong.  A healthy research culture has to begin by noticing that asymmetry, because almost every well-intentioned mechanism for ensuring ``responsible'' AI use redistributes risk toward those with the least power to absorb it.

The clearest example is automated detection. \cite{liang2023detectors} found that seven widely used GPT detectors achieved near-perfect accuracy on writing by native English speakers, yet misclassified over half of the human-written text by non-native speakers as AI-generated, yielding an average false positive rate of 61\%. Virtually all non-native essays (97.8\%) were flagged by at least one detector. Because non-native writers tend to produce lower-perplexity text with reduced lexical variability, it is difficult for automated detectors to distinguish between this and AI-generated text. When journals or research groups adopt AI detection for enforcement, they systematically risk falsely flagging non-native speakers or researchers who rely on AI as a genuine assistive technology.  These are often the very individuals who have the most to gain from AI assistance and the least institutional standing to contest a false accusation.


The research group is the smallest unit capable of mitigating these inequities by establishing shared AI norms to collectively absorb risks imposed by an uneven external environment. By ensuring that disclosing AI use for accessibility or language support is unequivocally safe, a principal investigator creates an environment where people can honestly describe use of AI tools (see Section~\ref{sec:transparency}). Groups must also address a critical second-order effect. As AI accelerates raw output, informal productivity metrics like papers per year, code shipped, and grants submitted increase very quickly. Scientists with high AI leverage implicitly establish a new, inflated baseline. Researchers are then evaluated against this standard, despite lacking the foundational domain expertise needed to safely oversee AI outputs. A research group that fails to explicitly address this dynamic will enable it by default.

\subsection{Meaningful Human Ownership}

Meaningful human ownership refers to maintaining an active intellectual role in the research process, and is a key element in editorial guidelines for astronomy journals \citep[e.g.][]{2023BAAS...55..016V, 2026NatAs..10..467.}. While AI tools can assist with nearly every stage of the research process, one choice should be made with intention and care:  the choice to delegate to AI the key decisions that shape the direction, execution, and interpretation of their work. Human ownership is a higher bar than simply spot-checking and approving an AI-generated output; it requires understanding how a result was produced, being able to explain and defend it, and exercising judgment about whether it is correct and appropriate.

\subsection{Retained Human Expertise}
Preserving the knowledge, skills, and scientific judgment required to conduct research independently is a high priority for scientist development \citep[c.f.~ the discussion in Section 2 of][which lays out a compelling argument that ``people are always the ends, not merely the means.'']{2026arXiv260210181H}. While AI tools can reduce the effort needed to perform many tasks, there are cognitive risks associated with offloading many of our intellectual tasks to artificial systems \citep[e.g.,][]{2025arXiv250608872K}.  Language AI tools can also reduce opportunities to develop and maintain expertise if researchers become overly dependent on automated assistance. Laboratories that prioritize scientist development therefore consider not only whether AI can perform a task, but whether delegating that task diminishes the expertise the group wishes to cultivate. 

\subsection{Policy Considerations}
Laboratories that prioritize scientist development will seek to avoid forms of AI use that transform researchers into passive reviewers of machine-generated work. Instead, AI should be used in ways that support learning, exploration, and productivity while preserving the researcher's role as the primary driver of scientific inquiry. A useful guiding question is: \textit{If this work succeeds or fails, do I genuinely understand it well enough to claim responsibility for the outcome, and to ensure its reproducibility?}  

AI also generates invisible labor. Every hallucinated citation, subtly flawed derivation, or plausible but incorrect line of code requires manual verification. This unglamorous, unrewarded work inevitably flows along existing academic fault lines, and may fall disproportionately on junior researchers and women, who already absorb the majority of non-promotable service tasks \citep{babcock2017}. A research culture that celebrates AI-driven productivity without explicitly allocating the verification burden simply shifts visible labor into invisible domains.  This can be mitigated by guidelines stating that this responsibility must not be delegated, and that the individual who deploys an AI tool assumes full ownership of verifying its output.

Some activities may be intentionally retained as human responsibilities because the educational value of performing the work exceeds the efficiency gained through automation. The goal is not to reject AI assistance, but we should strive to ensure that critical scientific knowledge is built within the research group rather than being outsourced to a tool. Additionally, AI assistance may be incorporated into training plans to mitigate risks of knowledge erosion.

\bigskip

\section{Scientific Integrity}
\label{sec:integrity}
Scientific integrity is a priority for every research laboratory, but it is \textit{the} defining priority of the \qualityaboveall research laboratories, who most value producing durable science.  With their prioritization of the trustworthiness of their output, this team focuses on adopting AI that supports their core question: \textbf{Can I trust the process and result?}

Scientific integrity requires more than accurate outputs; it requires confidence that the research process itself is trustworthy. Human evaluation, transparency and disclosure, and academic integrity provide complementary mechanisms to ensure that AI-assisted research remains rigorous, reproducible, and accountable.

\subsection{Human Evaluation}  

Generative AI systems can produce outputs that appear plausible, authoritative, and well-reasoned even when they contain various forms of hallucination: factual errors, fabricated citations, incorrect analyses, or flawed assumptions. Scientific integrity requires that researchers remain responsible for evaluating both the process and the results of AI-assisted work.  The ``right answer'' is simply not enough; scientific integrity requires there to be enough information for other researchers to evaluate what has been done. 

Human evaluation therefore extends beyond simple fact-checking. Researchers should consider the risks associated with a given application, the consequences of failure, and the degree to which AI outputs can be independently verified. High-consequence uses of AI may require formal evaluation procedures, benchmark datasets, or carefully curated ``golden samples'' against which outputs can be compared \citep[as in, e.g.,][]{2024arXiv240520389W, Hyk2025}. In other contexts, evaluation can take the form of human-led code review, comparison with the literature, replication of analyses, or consultation with other experts. Regardless of the specific method, the goal remains the same: to ensure that AI-assisted results meet the standards of reliability and evidence expected of scientific work.

\subsection{Transparency \& Disclosure}
\label{sec:transparency}

Transparency and disclosure involve documenting how AI tools contribute to the development of scholarly work. Rather than focusing on whether a tool was used, transparency emphasizes the process by which research and writing were produced. The relevant question is not “Did AI contribute to this work?” but “How did AI participate in the development of this work?” This approach recognizes that research outputs rarely emerge from a single author working in isolation. Instead, they are the product of what writing scholars sometimes describe as a “writing ecology”: a network of people, tools, feedback, and iterative revisions that shape the final result. AI systems are increasingly becoming one component of this broader scholarly ecosystem, alongside tools such as citation managers, statistical software, and code libraries.

Effective disclosure therefore documents the role that AI played within this ecosystem. A useful disclosure statement might describe what tool and tool version were used, what the tool was used for, what outputs were accepted or rejected, how the author modified or refined those outputs, and/or what steps were taken to verify accuracy or quality. Such statements help demonstrate that the human author remained an active participant and is ultimately responsible for the intellectual process. Transparency also supports reproducibility, a foundational principle of scientific research. In research workflows that incorporate AI, particularly for coding, analysis, or large-scale text processing, documenting the role of tools, prompts, and validation steps can make it easier for collaborators and future researchers to understand how results were produced. Reproducibility may also require archiving or publishing AI-generated code, workflows, or other computational outputs.

\subsection{Academic Integrity}
Expectations for disclosure will vary depending on the context. Journals, funding agencies, institutions, and collaborations are likely to have their own policies governing AI use \citep[e.g.][]{AandA_AI_Statement_2024, MNRASInstructions,  STScI2026HSTGAIPolicy}. Teams should therefore ensure that their internal practices remain compatible with relevant external requirements.

\subsection{Policy Considerations}

Policies alone are rarely sufficient to shape behavior, and scientific integrity is a place where this effect can be especially pronounced. In practice, teams operate according to shared social expectations about what forms of tool use are acceptable, expected, or discouraged. Without a common understanding of how AI tools should be used and discussed, individuals may default to either quiet experimentation, complete avoidance, or unilateral adoption simply because it appears that others are doing the same.

A clear culture of transparency helps avoid this dynamic. When researchers openly discuss their workflows, including how AI tools assisted in drafting, coding, or exploring ideas, AI becomes part of the normal research toolkit rather than something that must be hidden or justified. Transparency also helps preserve trust within collaborative research environments. Teams must be able to understand how work was produced, evaluate the reliability of results, and build on one another’s contributions. Disclosure practices, such as documenting process and provenance, support these goals while allowing researchers to benefit from new tools.

Importantly, transparency should not be framed as surveillance or enforcement. Instead, it should be understood as a continuation of familiar scholarly practices: documenting methods, citing sources, and acknowledging the tools that shaped the research process.

These can only function if disclosure is genuinely safe, prioritizing the guidelines in Section \ref{sec:culture}. A group that treats AI disclosure as evidence of integrity, rather than grounds for suspicion, makes those norms accessible to the people who need them most. 

At the field level, any institutional rule changes governing grants, peer review, or preprint servers must be developed with direct input from the early-career researchers they disproportionately impact, not merely imposed upon them.  A public capability layer including open astronomy-adapted models, shared benchmarks, and documented workflows is another way to address this.  When these public tools are built and released by the institutions and survey collaborations that can afford to make and maintain them, the entire community benefits.

\bigskip

\section{Data Governance}
\label{sec:data}
Data governance is the top priority of \stewardship laboratories, who see their data, software, and documentation as core deliverables to the rest of the astronomy community.  They consider their AI usage with the key question: \textbf{Is this approach secure, compliant, and responsible?}

Three pillars of data governance are important to consider when defining an appropriate Language AI policy for a research laboratory: security, privacy, and data integrity. Together, these address the risks associated with what information enters an AI system, who may gain access to it, and whether the outputs returned by the system can be trusted.

\subsection{Security}
Security refers to access and system protection. Research groups should consider who has access to the AI system, where submitted data are stored, and whether the system and its associated data are protected from malicious or unauthorized access. A tool that is convenient for low-risk work may be inappropriate for unpublished data, proprietary software, confidential collaborations, or institutional information.

\subsection{Privacy}
Privacy concerns whether the data provided to an AI agent are appropriate to share beyond the group. Researchers should consider the downstream consequences if those data became public, were retained by the service provider, or were used for model training or other purposes. Inputs that include personally identifiable information, personnel records, student information, collaborator communications, confidential institutional data, or sensitive unpublished results should be treated with particular care. Privacy policies and data-use terms should be understood before the tool is used, not after sensitive information has already been submitted.

\subsection{Data Integrity}
Data integrity concerns both how information is interpreted by an AI system and whether the resulting output is reliable. The data sent to an agent may not be processed in the way the user expects: it may be segmented, summarized, transformed, or combined with other information in ways that are not visible to the researcher. Likewise, the output returned by the agent may contain hallucinations, fabrications, incorrect citations, or misleading interpretations. AI-generated responses should therefore be used as guides, sources of ideas, or starting points for analysis, not as definitive scientific results. Important claims, code, citations, and analyses should be cross-referenced, tested, and independently verified before use.

\subsection{Policy Considerations}

Different AI deployment models carry different levels of risk. Free or public AI tools are typically the highest risk because they may provide limited control over data retention, model training, and data residency. Enterprise AI tools may offer stronger privacy protections, but their terms, defaults, and opt-in or opt-out requirements should be carefully reviewed. Self-hosted or local models generally provide the greatest control over data handling, but determining whether a system is truly local, what data it logs, and what services it contacts can be technically complex. A research group's policy should therefore specify which types of data may be used with which classes of AI tools and under what conditions.  

The Open Worldwide Application Security Project (OWASP) is a nonprofit organization and community that provides free resources to improve the security of software.  For a more thorough exploration of the issues of Language AI security, see their report on Generative AI Security \citep{owasp_llm_top10_2025}, which provides a deeper exploration into the vulnerabilities of Language AI and how to adequately address them.

\bigskip
\section{Sample Policy Guidelines}
\label{sec:sample}

Each research lab should engage with Language AI tools in ways that support its research culture and scientific priorities.   In this section, we discuss how and where each archetype lab will engage with Language AI tools, and expand on the caveats and checkpoints each lab is likely to prioritize when using AI tools.  Table \ref{table:sample_policies} provides an overview summary of how each group might embrace or forbid specific tools and use cases in their laboratory policy and Appendix  \ref{sec:example} provides a concrete example of an AI policy currently in use by several of the authors.

The \gettherefast research lab prioritizes maximizing the efficiency and speed of scientific discovery.  This team adopts AI broadly and aggressively, with restraint reserved only for high-risk approaches.  They prioritize Language AI applications where they reduce bottlenecks, and routinely use AI for code generation, debugging, literature review, data analysis, brainstorming, drafting text, and automating administrative tasks.  The \gettherefast lab archetype will upload sensitive data to Language AI only when the benefits outweigh the risks.  While concerns regarding validation, transparency, and security are not ignored, they are generally weighed against the potential cost of moving more slowly than competing groups. In this model, AI functions as a force multiplier: a tool for expanding research capacity, accelerating iteration, and maximizing the pace of scientific discovery.

The \apprentice research lab prioritizes developing skilled, independent scientists through practice and mentorship.  This team adopts AI only when it supports the educational process, but avoids shortcuts that bypass the acquisition of expertise.  Meeting summarization is one of the first Language AI tools that this team will adopt, as the tool preserves team knowledge and improves documentation without interrupting the mentorship of trainees.  AI tools may be used as tutors, discussion partners, or educational aids—for example, to explain concepts, suggest debugging strategies, summarize literature for discussion, or provide feedback on writing. However, researchers are generally expected to attempt tasks independently before consulting AI systems, and AI-generated outputs are treated as opportunities for learning rather than authoritative answers.  This lab archetype is most likely to have a Language AI policy that is not a list of set rules, but adapts to the seniority and expertise of the user \textemdash{} while students might be discouraged from using AI code generation until they have developed appropriate fluency, more seasoned researchers might openly and heavily rely on the tool. Uses of AI that risk replacing core intellectual activities, such as drafting papers, automating research decisions, or obscuring individual ownership of the work, are discouraged. In this model, AI serves not as a substitute for expertise, but as a tool for cultivating it.

The \qualityaboveall research lab prioritizes producing results that are rigorous, reproducible, and trustworthy.  This team adopts AI only where it provides support in improving the quality of the research group's science.  This laboratory approaches Language AI cautiously, viewing it as a potentially useful tool whose outputs must be independently verified before they can be trusted.  Members of this lab archetype will tend to adopt Language AI first for code debugging and writing assistance, along with other administrative tasks.  This group adopts tools with substantial restrictions on how and where tools can be used, oversight at every step, and/or human validation requirements to confirm that the output is correct (these restrictions and caveats are denoted with a $\triangle$ symbol in table \ref{table:sample_policies}).  This lab archetype approaches Language AI with cautious optimism, but AI contributions are almost always subject to careful human evaluation.

The \stewardship research lab prioritizes the responsible management of data, software, institutional resources, and public trust. This team evaluates Language AI first through the lens of security, privacy, compliance, and data integrity. AI tools may be adopted when they can be used within approved environments, with clear safeguards governing what information may be shared, stored, or processed.  This lab will prefer self-hosted models over public models and will prioritize confidentiality and data security over speed.  Public cloud tools, autonomous agents, and uploads of sensitive or proprietary data are treated cautiously or prohibited when they create unacceptable risks, but this lab archetype may adopt a more open or permissive framework in cases where the AI tool promises to make data products and software more open and accessible to community use.

\begin{deluxetable*}{lcccc}
\tablecaption{Illustrative attitudes toward different classes of AI tools among the four laboratory archetypes.\label{table:sample_policies}}
\tablehead{
\colhead{AI Capability} &
\colhead{\gettherefast} &
\colhead{\apprentice} &
\colhead{\qualityaboveall} &
\colhead{\stewardship}
}
\startdata
Code generation            & \yes & \maybe & \maybe & \maybe \\
Code debugging             & \yes & \maybe & \yes & \maybe \\
Literature summarization   & \yes & \maybe & \maybe & \maybe \\
Writing assistance         & \yes & \yes & \yes & \yes \\
Paper/proposal drafting    & \yes & \no & \maybe & \maybe \\
Meeting summarization      & \yes & \yes & \yes & \maybe \\
Brainstorming and ideation & \yes & \maybe & \maybe & \maybe \\
Data-analysis assistance   & \yes & \maybe & \maybe & \maybe \\
Autonomous AI agents       & \yes & \no & \maybe & \no \\
Public cloud AI services   & \yes & \maybe & \maybe & \no \\
Local/self-hosted models   & \maybe & \maybe & \maybe & \yes \\
Sensitive data as AI inputs     & \maybe & \no & \no & \no \\
\enddata

\tablenotetext{}{%
\begin{tabular}{@{}lcl@{}}
\yes   & = & generally adopted; \\
\maybe & = & adopted with restrictions, oversight, or validation requirements; \\
\no    & = & generally discouraged or prohibited.\\
\end{tabular}\\
Section \ref{sec:sample} provides more detail about how each archetype lab group might define the use cases that are appropriate for Language AI and the guardrails against improper use.}
\end{deluxetable*}

\bigskip
\section{Conclusion}
\label{sec:conclusion}

The version of astronomy that exists in five years will not be shaped by any single group's policy, but by the aggregate of choices being made today. The astronomers who will pose the next generation of questions are being trained in groups like this one right now. The choices made today about who these tools are built to serve will determine who those astronomers are.

Every research group is faced with the same fundamental question when adopting Language AI: what work should be delegated to machines, and what work should remain part of the educational process of becoming a scientist?\\

\textbf{There is no universal answer.} \\

\noindent The appropriate balance depends on the laboratory's values, whether it is to be the first to a new result, to develop mentees with deep expertise, to work together as a coherent and supportive research team, or to produce the highest quality research.  Language AI creates a new tension between delegation and education: every task assigned to a machine is a task that may no longer contribute to the development of human expertise. Each laboratory must therefore determine for itself how to balance these competing goals and develop policies that reflect its priorities and values.

This is why we believe that effective AI policies don't begin with a generic prescription that lists what is permitted and what is prohibited. Instead, they should support the research values that the group prioritizes \textemdash{} human development, scientific integrity, data governance, and research productivity are all legitimate priorities, and different laboratories will weight them differently.

A good policy should help a research team identify when AI use is appropriate, when human expertise must remain central, and when delegation would compromise learning, accountability, or trust.

Language AI will continue to change the practice of research, and AI policies should be thoughtful guidelines aligned with a team's priorities. Teams and technologies both evolve, and so should AI policies. They should be living documents that are revisited regularly, informed by experience, and responsive to the needs of the research team. Every laboratory's AI policy comes with some assembly required. 

\section*{AI Disclosure}
Generative AI tools, including ChatGPT and Claude were used during the development of this manuscript. The authors employed these tools to brainstorm organizational structures, explore alternative framings, suggest titles and section headings, refine prose, identify redundancies, generate illustrative examples, and improve clarity. AI-generated content was reviewed, evaluated, modified, and, in many cases, rejected or substantially rewritten by the authors. The authors retained responsibility for the intellectual content, arguments, interpretations, and conclusions presented in this work.

\acknowledgements{}  
The conversations that led to the development of this white paper were facilitated by the \href{https://www.stsci.edu/contents/events/stsci/2026/march/language-ai-in-the-space-sciences}{Language AI in the Space Sciences Workshop}, which was co-organized by the Data Science Mission Office (DSMO) at the Space Telescope Science Institute (STScI), the European Space Agency (ESA), and the Astrophysics Data System (ADS).  This workshop was hosted March 9-12, 2026, at the Space Telescope Science Institute in Baltimore, MD, USA. We acknowledge the contributions of J.E.G.~Peek to this work.  Ana Maria Delgado acknowledges support from NSF grant number 2307070.\vspace{5em}

\clearpage
\bibliographystyle{apj}
\bibliography{references}

@ARTICLE{2026NatAs..10..467.,
        author = {{Nature Astronomy}},
        title = "{Towards a coordinated approach to LLMs in astronomy}",
      journal = {Nature Astronomy},
         year = 2026,
        month = apr,
       volume = {10},
        pages = {467-467},
          doi = {10.1038/s41550-026-02858-x},
       adsurl = {https://ui.adsabs.harvard.edu/abs/2026NatAs..10..467.}
}

@misc{STScI2026HSTGAIPolicy,
  author       = {{Space Telescope Science Institute}},
  title        = {HST Guidelines and Checklist for Phase I Proposal Preparation: Use of Generative Artificial Intelligence (GAI) Technology},
  year         = {2026},
  howpublished = {\url{https://hst-docs.stsci.edu/hsp/hubble-space-telescope-call-for-proposals-for-cycle-34/hst-guidelines-and-checklist-for-phase-i-proposal-preparation}},
  note         = {{Hubble} Space Telescope Cycle 34 Call for Proposals. Accessed 2026-07-23},
}

@online{MNRASInstructions,
  author       = {{Monthly Notices of the Royal Astronomical Society}},
  title        = {Instructions to Authors},
  year         = {2026},
  url          = {https://academic.oup.com/mnras/pages/General_Instructions},
  urldate      = {2026-07-23},
  organization = {Oxford University Press}
}

@ARTICLE{2026arXiv260329039H,
       author = {{Huang}, Chun},
        title = "{AI Cosplaying as Astrophysicists: A Controlled Synthetic-Agent Study of AI-Assisted Astrophysical Research Workflows}",
      journal = {arXiv e-prints},
         year = 2026,
        month = mar,
          eid = {arXiv:2603.29039},
        pages = {arXiv:2603.29039},
          doi = {10.48550/arXiv.2603.29039},
archivePrefix = {arXiv},
       eprint = {2603.29039},
 primaryClass = {astro-ph.IM},
       adsurl = {https://ui.adsabs.harvard.edu/abs/2026arXiv260329039H}
}

@ARTICLE{2025arXiv250520538J,
       author = {{Joseph}, Sebastian Antony and {Murtaza Husain}, Syed and {Offner}, Stella S.~R. and {Juneau}, St{\'e}phanie and {Torrey}, Paul and {Bolton}, Adam S. and {Farias}, Juan P. and {Gaffney}, Niall and {Durrett}, Greg and {Jessy Li}, Junyi},
        title = "{AstroVisBench: A Code Benchmark for Scientific Computing and Visualization in Astronomy}",
      journal = {arXiv e-prints},
         year = 2025,
        month = may,
          eid = {arXiv:2505.20538},
        pages = {arXiv:2505.20538},
          doi = {10.48550/arXiv.2505.20538},
archivePrefix = {arXiv},
       eprint = {2505.20538},
 primaryClass = {cs.CL},
       adsurl = {https://ui.adsabs.harvard.edu/abs/2025arXiv250520538J}}

@ARTICLE{2026arXiv260318145S,
       author = {{Schuster}, Nico and {Salcedo}, Andr{\'e}s N. and {Bouchard}, Simon and {Frei}, Dennis and {Pisani}, Alice and {Bautista}, Julian E. and {Zoubian}, Julien and {Escoffier}, Stephanie and {Liu}, Wei and {Valogiannis}, Georgios and {Zarrouk}, Pauline},
        title = "{Setting SAIL: Leveraging Scientist-AI-Loops for Rigorous Visualization Tools}",
      journal = {arXiv e-prints},
         year = 2026,
        month = mar,
          eid = {arXiv:2603.18145},
        pages = {arXiv:2603.18145},
          doi = {10.48550/arXiv.2603.18145},
archivePrefix = {arXiv},
       eprint = {2603.18145},
 primaryClass = {astro-ph.IM},
       adsurl = {https://ui.adsabs.harvard.edu/abs/2026arXiv260318145S}
}

@ARTICLE{moss+aicosmologist,
       author = {{Moss}, Adam},
        title = "{The AI Cosmologist I: An Agentic System for Automated Data Analysis}",
      journal = {arXiv e-prints},
         year = 2025,
        month = apr,
          eid = {arXiv:2504.03424},
        pages = {arXiv:2504.03424},
          doi = {10.48550/arXiv.2504.03424},
archivePrefix = {arXiv},
       eprint = {2504.03424},
 primaryClass = {astro-ph.IM},
       adsurl = {https://ui.adsabs.harvard.edu/abs/2025arXiv250403424M}
}

@misc{AandA_AI_Statement_2024,
  author       = {{Astronomy \& Astrophysics}},
  title        = {A\&A Publishes Statement on the Use of AI-Assisted Technologies},
  year         = {2024},
  howpublished = {\url{https://www.aanda.org/component/content/article/11-news/3200-a-a-publishes-statement-on-the-use-of-ai-assisted-technologies}},
  note         = {Accessed: 2026-06-19}
}

@techreport{owasp_llm_top10_2025,
  author       = {{OWASP}},
  title        = {{OWASP Top 10 for LLM Applications 2025}},
  institution  = {OWASP Foundation},
  year         = {2024},
  month        = nov,
  type         = {Version 2025},
  url          = {https://owasp.org/www-project-top-10-for-large-language-model-applications/assets/PDF/OWASP-Top-10-for-LLMs-v2025.pdf},
  note         = {Released November 18, 2024; accessed 2026-06-12}
}

@misc{copilot,
    title={Research: Quantifying GitHub Copilot’s impact on code quality},
    author={Rodriguez, Mario},
    url={https://github.blog/2023-10-10-research-quantifying-github-copilots-impact-on-code-quality/},
    year={2023}
}

@article{codegen,
  title={CodeGen2: Lessons for Training LLMs on Programming and Natural Languages},
  author={Nijkamp, Erik and Hayashi, Hiroaki and Xiong, Caiming and Savarese, Silvio and Zhou, Yingbo},
  journal={ICLR},
  year={2023}}

@misc{GPT3,
      title={Language Models are Few-Shot Learners}, 
      author={Tom B. Brown and Benjamin Mann and Nick Ryder and Melanie Subbiah and Jared Kaplan and Prafulla Dhariwal and Arvind Neelakantan and Pranav Shyam and Girish Sastry and Amanda Askell and Sandhini Agarwal and Ariel Herbert-Voss and Gretchen Krueger and Tom Henighan and Rewon Child and Aditya Ramesh and Daniel M. Ziegler and Jeffrey Wu and Clemens Winter and Christopher Hesse and Mark Chen and Eric Sigler and Mateusz Litwin and Scott Gray and Benjamin Chess and Jack Clark and Christopher Berner and Sam McCandlish and Alec Radford and Ilya Sutskever and Dario Amodei},
      year={2020},
      eprint={2005.14165},
      archivePrefix={arXiv},
      primaryClass={cs.CL}
}

@inproceedings{vaswani2017attention,
	title        = {{Attention is All You Need}},
	author       = {Vaswani, Ashish and Shazeer, Noam and Parmar, Niki and Uszkoreit, Jakob and Jones, Llion and Gomez, Aidan N and Kaiser, {\L}ukasz and Polosukhin, Illia},
	year         = 2017,
	booktitle    = nips,
	url          = {https://arxiv.org/abs/1706.03762},
}

@book{decadal2020,
	author = {{NASEM}},
	number = {https://doi.org/10.17226/26141.},
	publisher = {Washington, DC: The Washington, DC: National Academies Press.},
	title = {{Pathways to Discovery in Astronomy and Astrophysics for the 2020s.}},
	year = 2021}

@article{wang2025effect,
  title={The effect of ChatGPT on students’ learning performance, learning perception, and higher-order thinking: insights from a meta-analysis},
  author={Wang, Jin and Fan, Wenxiang},
  journal={Humanities and Social Sciences Communications},
  volume={12},
  number={1},
  pages={1--21},
  year={2025},
  publisher={Palgrave}
}

@ARTICLE{2025arXiv250608872K,
       author = {{Kosmyna}, Nataliya and {Hauptmann}, Eugene and {Yuan}, Ye Tong and {Situ}, Jessica and {Liao}, Xian-Hao and {Beresnitzky}, Ashly Vivian and {Braunstein}, Iris and {Maes}, Pattie},
        title = "{Your Brain on ChatGPT: Accumulation of Cognitive Debt when Using an AI Assistant for Essay Writing Task}",
      journal = {arXiv e-prints},
         year = 2025,
        month = jun,
          eid = {arXiv:2506.08872},
        pages = {arXiv:2506.08872},
          doi = {10.48550/arXiv.2506.08872},
archivePrefix = {arXiv},
       eprint = {2506.08872},
 primaryClass = {cs.AI},
       adsurl = {https://ui.adsabs.harvard.edu/abs/2025arXiv250608872K}
}

@ARTICLE{2023BAAS...55..016V,
       author = {{Vishniac}, Ethan T.},
        title = "{Editorial: On the Use of Chatbots in Writing Scientific Manuscripts}",
      journal = {\baas},
         year = 2023,
        month = mar,
       volume = {55},
       number = {1},
          eid = {016},
        pages = {016},
          doi = {10.3847/25c2cfeb.c3619710},
       adsurl = {https://ui.adsabs.harvard.edu/abs/2023BAAS...55..016V}
}

@ARTICLE{2024arXiv240520389W,
       author = {{Wu}, John F. and {Hyk}, Alina and {McCormick}, Kiera and {Ye}, Christine and {Astarita}, Simone and {Baral}, Elina and {Ciuca}, Jo and {Cranney}, Jesse and {Field}, Anjalie and {Iyer}, Kartheik and {Koehn}, Philipp and {Kotler}, Jenn and {Kruk}, Sandor and {Ntampaka}, Michelle and {O'Neill}, Charles and {Peek}, Joshua E.~G. and {Sharma}, Sanjib and {Yunus}, Mikaeel},
        title = "{Designing an Evaluation Framework for Large Language Models in Astronomy Research}",
      journal = {arXiv e-prints},
         year = 2024,
        month = may,
          eid = {arXiv:2405.20389},
        pages = {arXiv:2405.20389},
          doi = {10.48550/arXiv.2405.20389},
archivePrefix = {arXiv},
       eprint = {2405.20389},
 primaryClass = {astro-ph.IM},
       adsurl = {https://ui.adsabs.harvard.edu/abs/2024arXiv240520389W}
}

@ARTICLE{2026arXiv260210181H,
       author = {{Hogg}, David W.},
        title = "{Why do we do astrophysics?}",
      journal = {arXiv e-prints},
         year = 2026,
        month = feb,
          eid = {arXiv:2602.10181},
        pages = {arXiv:2602.10181},
          doi = {10.48550/arXiv.2602.10181},
archivePrefix = {arXiv},
       eprint = {2602.10181},
 primaryClass = {astro-ph.IM},
       adsurl = {https://ui.adsabs.harvard.edu/abs/2026arXiv260210181H}
}

@misc{AASCodeEthics,
  author       = {{American Astronomical Society}},
  title        = {AAS Code of Ethics},
  year         = {2023},
  howpublished = {\url{https://aas.org/policies/ethics}},
  note         = {Last updated 2 August 2023. Accessed 11 March 2026}
}

@article{mehr2020write,
  title={How to... write a lab handbook.},
  author={Mehr, Samuel},
  journal={Biologist},
  volume={67},
  number={2},
  year={2020}
}

@article{noy2023experimental,
author = {Shakked Noy  and Whitney Zhang },
title = {Experimental evidence on the productivity effects of generative artificial intelligence},
journal = {Science},
volume = {381},
number = {6654},
pages = {187-192},
year = {2023},
doi = {10.1126/science.adh2586},
URL = {https://www.science.org/doi/abs/10.1126/science.adh2586},
eprint = {https://www.science.org/doi/pdf/10.1126/science.adh2586}}

@ARTICLE{liang2023detectors,
       author = {{Liang}, Weixin and {Yuksekgonul}, Mert and {Mao}, Yining and {Wu}, Eric and {Zou}, James},
        title = "{GPT detectors are biased against non-native English writers}",
      journal = {arXiv e-prints},
         year = 2023,
        month = apr,
          eid = {arXiv:2304.02819},
        pages = {arXiv:2304.02819},
          doi = {10.48550/arXiv.2304.02819},
archivePrefix = {arXiv},
       eprint = {2304.02819},
 primaryClass = {cs.CL},
       adsurl = {https://ui.adsabs.harvard.edu/abs/2023arXiv230402819L}
}

@article{delllacqua2023jagged,
author = {Dell’Acqua, Fabrizio and McFowland, Edward and Mollick, Ethan and Lifshitz, Hila and Kellogg, Katherine C. and Rajendran, Saran and Krayer, Lisa and Candelon, Fran\c{c}ois and Lakhani, Karim R.},
title = {Navigating the Jagged Technological Frontier: Field Experimental Evidence of the Effects of Artificial Intelligence on Knowledge Worker Productivity and Quality},
journal = {Organization Science},
volume = {37},
number = {2},
pages = {403-423},
year = {2023},
doi = {10.1287/orsc.2025.21838},
URL = { https://doi.org/10.1287/orsc.2025.21838},
eprint = { https://doi.org/10.1287/orsc.2025.21838}
}

@ARTICLE{caplar2017,
       author = {{Caplar}, Neven and {Tacchella}, Sandro and {Birrer}, Simon},
        title = "{Quantitative evaluation of gender bias in astronomical publications from citation counts}",
      journal = {Nature Astronomy},
         year = 2017,
        month = jun,
       volume = {1},
          eid = {0141},
        pages = {0141},
          doi = {10.1038/s41550-017-0141},
archivePrefix = {arXiv},
       eprint = {1610.08984},
 primaryClass = {astro-ph.IM},
       adsurl = {https://ui.adsabs.harvard.edu/abs/2017NatAs...1E.141C}
}

@article{babcock2017,
  title={Gender differences in accepting and receiving requests for tasks with low promotability},
  author={Babcock, Linda and Recalde, Maria P and Vesterlund, Lise and Weingart, Laurie},
  journal={American Economic Review},
  volume={107},
  number={3},
  pages={714--747},
  year={2017},
  publisher={American Economic Association}
}

@inproceedings{Hyk2025,
title={From Queries to Criteria: Understanding How Astronomers Evaluate {LLM}s},
author={Alina Hyk and Kiera McCormick and Mian Zhong and Ioana Ciuc{\u{a}} and Sanjib Sharma and John F Wu and J. E. G. Peek and Kartheik G. Iyer and Ziang Xiao and Anjalie Field},
booktitle={Second Conference on Language Modeling},
year={2025},
url={https://openreview.net/forum?id=ROtDZDUgvw}
}

@misc{Liu2026,
      title={AI Assistance Reduces Persistence and Hurts Independent Performance}, 
      author={Grace Liu and Brian Christian and Tsvetomira Dumbalska and Michiel A. Bakker and Rachit Dubey},
      year={2026},
      eprint={2604.04721},
      archivePrefix={arXiv},
      primaryClass={cs.AI},
      url={https://arxiv.org/abs/2604.04721}, 
}

\appendix{}

\section{Example AI Policy}
\label{sec:example}

The following AI policy was developed and adopted by the Chesapeake ML-Astro Research Group in early 2025, as Language AI models were becoming more prevalent, to be a set of guiding principles for the use of Language AI in the team's research, teaching, and scholarly communication. In many ways, this policy served as the starting point for the ideas explored throughout this white paper. We include it here not as a template to be copied verbatim, but as a concrete example of how one research group translated its values and priorities into practice. Readers may find that some elements resonate strongly with their own approach to research, while others reflect different assumptions about the role of AI in scientific work. As an exercise, we invite readers to consider which of the laboratory archetypes described in this paper most closely align with the philosophy embodied in the policy below.

\bigskip

Before using the output of AI productivity tools for your research (including both code and writing), you should be able to answer “yes” to the following questions.  

\textbf{Is this method adequately secure?}  Most free (and even some paid) Language AI tools recycle user inputs and generated outputs for future training.  Would you be okay with the query or answer being published on the front page of the New York Times?

\textbf{Would I stake my professional reputation on this output?}  The output should stand up to evaluation from peers/managers/professors the same as your own work.  Do not use Language AI to skip critical stages in your own education or professional development.

\textbf{Do I have the expertise to critically evaluate this output?}  AI tools may give false or biased results, and you need to understand the consequences of AI getting it wrong, and avoid scenarios where a wrong answer could negatively affect someone else’s life.  Do not use Language AI in a way that creates work for your peers, your manager, or your professor.

\textbf{Does this give me the appropriate level of ownership of the work?} Work that you submit should be consistent with your voice, your style, and your values.  

\textbf{Does this meet the academic integrity guidelines of the place where I’ll be submitting the work?}  Academic integrity and disclosure standards vary across industries, organizations, and classes.  You are responsible for finding and following GenAI guidelines when you submit work.

\section{Some Assembly Required}
\label{sec:blank}

Before writing a policy, ask each member of the research group to independently complete the diagram on the following page. Differences between profiles often reveal unstated assumptions about productivity, training, integrity, and governance that should be discussed before drafting a policy.  Comparing the resulting profiles can help identify areas of agreement, uncover differing assumptions about acceptable tradeoffs, and provide a concrete starting point for policy discussions. The goal is not to discover the ``correct'' profile, but to facilitate a discussion about the values that will ultimately shape a laboratory's AI policy.

\begin{figure*}
    \centering
    \vspace{10em}
    \includegraphics[width=0.9\linewidth]{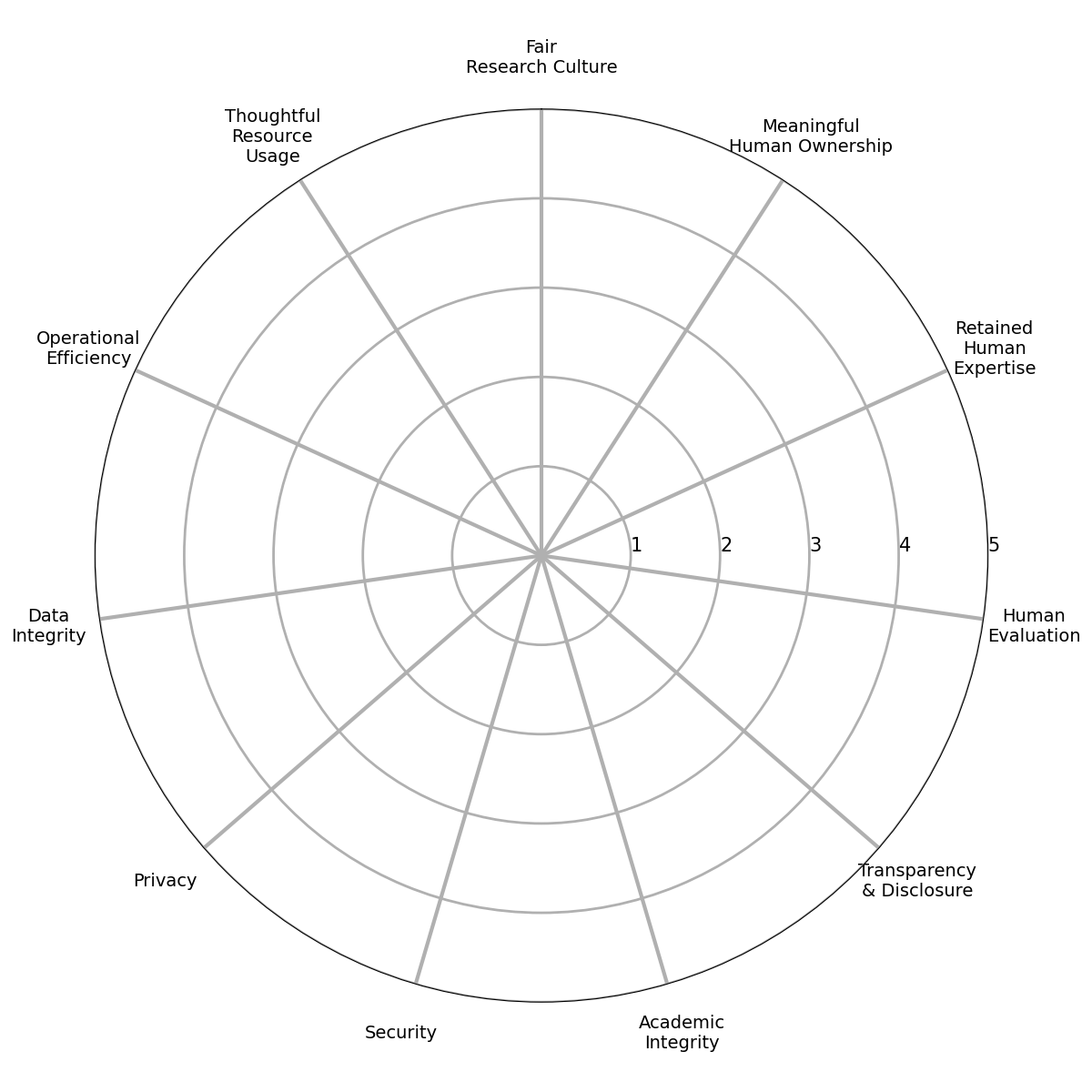}
    \vspace{10em}
    \caption{Blank laboratory priority profile template. Group assembly recommended. Some disagreement during assembly is normal and does not indicate a manufacturing defect.  Consensus sold separately.}
    \label{fig:blank}
\end{figure*}

\end{document}